# A METHODOLOGY FOR REALISTIC ESTIMATION OF THE AEROSOL IMPACT ON THE SOLAR POTENTIAL

Robert Blaga[1], Delia Calinoiu[2], Marius Paulescu[1,*]

[1]Faculty of Physics, West University of Timisoara, V. Parvan 4, 300223 Timisoara, Romania

[2]Department of Fundamental of Physics for Engineers, Politehnica University of Timisoara, V. Parvan 2, 300223 Timisoara, Romania

*Corresponding author: marius.paulescu@e-uvt.ro

ABSTRACT

The atmospheric aerosol loading may significantly influence the performance in solar power production. The impact can be very different both in space (even in short distance) and time (short-term fluctuations as well as long-term trend). Aiming to ensure a high degree of generality, this study is focused on the aerosol impact on collectable solar energy. Thus, the results are independent of solar plants' characteristics. A new methodology for estimating the average daily, monthly, and yearly losses in the solar potential due to aerosols is proposed. For highlighting the loss in the overall solar potential, a new ideal scenario is defined as a reference for the atmospheric aerosol background. A new equation for computing the solar potential loss is proposed to adjust for possible biases. In a departure from similar studies, the analysis relies on ground measurements (BSRN and AERONET), always more accurate than remotely sensed satellite data. The seldom discussed impact of aerosol type is considered. As a general conclusion, the monthly and yearly reductions of the solar potential due to aerosols are estimated at 12 locations spread around the globe, amounting to losses of the solar potential ranging from 0.6% to as high as 7.2%.

1. INTRODUCTION

Long-term assessment of photovoltaic (PV) power production is commonly based on an estimate of the solar potential as accurately as possible. In the design stage of a solar plant, a miscalculation in the estimation of the solar potential may significantly alter the revenue of the project. For example, Ref. [1], dealing with the return of invested capital in residential PV systems installed in Brazil, shows that the variation in payback years may achieve 30% when different solar resource data are used to evaluate residential projects. Different from PV systems, concentrating solar power (CSP) plants convert only the solar flux normally incident on the collector surface into heat and/or electricity. Therefore, the accurate estimation of direct-normal solar irradiance (DNI) is a requirement for estimating the solar power production for designing a CSP plant (see e.g. [2]).

There is a general agreement that the embedding of cloud shadow into the solar irradiance models represents the primary source of uncertainty in the solar potential estimation. However, the uncertainty may also emerge from the inability of accurately anticipating the atmospheric column content. Sometimes, limited knowledge of the atmospheric parameters may result in a dramatic difference between the estimated and measured solar power production. For instance, Ref. [3] showed that the estimated clear-sky energy yield by a CPV plant over an annual period can vary by up to 75% between the case when all available atmospheric parameters are used and the most basic case, when only the atmospheric air mass is considered. Such a difference in energy yields cases can increase the levelized cost of energy (LCOE) by up to 25% [3].

Aerosol is the most influential factor of atmospheric transmittance, but at the same time, the most volatile atmospheric constituent. Subsequently, the atmospheric aerosol loading influences differently in space (even in close locations) and time (short-term fluctuations as well as long-term trend) the performance in solar power production. Based on coupled aerosol-climate simulations that take into account the spatial and temporal patterns of natural and anthropogenic aerosols over the Euro-Mediterranean domain, the outstanding study [4] shows that the most affected area is Central Europe where sensitivity of PV production to aerosols is higher. For instance, in The Netherlands, the annual production loss due to aerosols ranges from no impact to 16%. The PV production loss in summer can even reach 20% over regions of Africa and Syria-Iraq [4]. There are various studies devoted to the aerosol impact in PV power production, some of them being acknowledged next [5-17].

The main causes of solar power loss due to aerosol are:

- *Soiling of PV modules* (e.g. [5]). Such a presence of the aerosol is visible to the naked eye. Without proper cleaning of PV modules, the accumulated dust effect on PV performance can be significant (e.g. 4% in Tokyo for a one-year exposure, increasing to astonishing 80% in Doha, Qatar for 140 days exposure [6]). There is a close relation between the atmospheric aerosol loading and the dust deposition. The recent study [7], dealing with this topic, analyzes the capabilities of two reanalysis products (MERRA-2 and CAMS) in capturing temporal cycles of aerosol over Australia and the impact in dust deposition.
- *Surface erosion of solar convertors due to the impact of dust particles,* contributing to the reduction of their performance (see e.g. [8]). The reduction in energy production, due to the increase in roughness of PV modules surface caused by dust, is only marginally investigated.
- *Decreasing the collectable solar energy due to aerosol extinction*, especially decreasing the direct-normal component, and increasing the diffuse component. Aerosol loading of the atmosphere may drastically reduce the PV power production. Severe events may reduce the PV power production at a level as high as that reported for soiling. A study on the 12-year average reduction of in-plane solar irradiance caused by the aerosol atmospheric loading in China was reported by [9]. The results show that over northern and eastern China—the most polluted regions—the annual average of in-plane solar irradiance decreases by up to 1.5 kWh/m$^2$ (35%) per day relative to non-polluting conditions [9]. In some regions of the world, the aerosol effect on solar radiation is comparable to that of clouds. For instance, clouds over the northern Arabian Peninsula in winter and desert dust over the central and southern Arabian Peninsula in summer are the major factors driving the variability of diffuse solar irradiance [10].

The atmospheric aerosol load and the aerosol nature (see e.g. [11] for aerosol classification) may equally influence the PV power production. Aiming to achieve accurate scenarios and improve the grid integration, both amount and type of aerosol should be taken into account in PV power evaluation [12]. There are two major sources of the atmospheric aerosol that can spread over large areas and can impact the collectable solar energy and PV power production: desert dust, e.g. [13] and biomass burning, e.g [14]. Even if many studies are focused on the effect of aerosol on PV power production, only a few are devoted to the effect of pollutant type on the PV performance [15]. Anthropogenic haze, caused at least in parts by forest and agricultural land clearing fires in Sumatra (Indonesia), is occasionally causing air quality issues in Singapore, located 150–300 km east of the majority of these "hot spots". An assessment of a major haze event in June 2013 in Singapore (caused by fires in Sumatra, Indonesia at 150-300 km) caused a loss of PV power production in the range of 15–25% [16]. The study reported in [17] discusses the PV power production as function of atmospheric aerosols (dust) in the semi-arid African region of Sahel for clear-sky-days, highlighting a reduction in daily global solar irradiation and PV power production with 13% and 14%, respectively. During extreme events, such as dust storms, PV power production can be reduced by half.

Without a doubt, there are numerous studies focused on the aerosol impact on PV power production. The reported results create a picture of the possible energy losses caused by aerosols. Still, this picture is most often accompanied by a certain degree of subjectivism, being more or less influenced by the particularities of solar plants subject to investigation. Differently, this study is focused on the aerosol impact on the collectable solar energy. The results are independent of solar plant characteristics, thus preserving a high degree of generality and applicability.

This study introduces a new methodology to estimate the average daily, monthly and yearly losses in the solar potential due to aerosols. In a departure from similar studies, the analysis relies on ground measurements—always more accurate than remotely sensed satellite data, which are most commonly used in solar potential studies. Compared to remotely sensed data, measurements from ground stations also have higher spatial granularity (pointlike vs. km$^2$ sized grid) and temporal resolution (minute vs. hourly and higher). The downside is the limited number of locations where professional monitoring stations are located. In this work, data from two public repositories are exploited, providing aerosol and solar data, respectively. A subset of locations is included in both networks, meaning that both solar irradiance and aerosol optical depth (AOD) are recorded with high quality at the same location, which is vital for such a study. Only clear-sky days are retained in order to single out the impact of aerosols on the collectable solar energy.

On brief, the novelties reported in this study are: (i) a new *ideal* scenario is defined as a reference for the atmospheric aerosol background; (ii) a new equation for computing the solar potential loss is proposed to adjust for possible biases; (iii) the seldom discussed impact of aerosol type is considered, (iv) the exclusive reliance on ground measurements.

The paper is structured as follows. In Sec. 2 the procedure for building the database is described, and the algorithm for assessing losses in the solar potential is presented. In Sec. 3 the aerosol-induced losses at various locations on daily, monthly, and yearly timescale are computed. The impact of the aerosol type is also studied. Results are summarized and further research paths are explored in the concluding section.

## 2. DATABASE AND METHODS

### 2.1. Datasets

The dataset used in this study consists of ground-based measurements from two public archives: the Baseline Surface Radiation Network - BSRN [18] and the Aerosol Robotic Network - AERONET [19]. BSRN is comprised of 71 past and present stations which measure the solar irradiance components at 1- or 3-minute resolution. AERONET is a network comprised of more than 600 stations that record aerosol optical, microphysical and radiative properties, generally at 15-minute resolution. There are several versions of the AERONET data which are available. The level 1.5 data is automatically screened for the presence of clouds, while a further manual inspection is performed to obtain the level 2 data.

This study was conducted on data recorded during the period 2010-2020, at locations which are included in both AERONET and BSRN. The stations are not always exactly collocated. The condition that the separation between the emplacement of the AERONET and BSRN stations be below 0.1° in both latitude and longitude was imposed during selection. In order to fully valorize the measured data, the single scattering albedo was retrieved from the level 1.5 AERONET data. For all other parameters, the highest quality data are used (level 2, version 3). The BSRN data was quality controlled using the algorithm proposed in [20], which is based on the BRSN recommendations [21, 22]. The algorithm is described in detail in [23]. Data lines with missing values for any of the relevant parameters were removed. A screening for clear-sky conditions was performed on the BSRN data. The state of the sky was classified using the quantifier $I$ developed by the authors in 2019 [24]. The indicator $I$ is based on the cumulative distribution function of clearness index increments. $I$ requires only the knowledge of the global solar irradiance and extraterrestrial horizontal irradiance A threshold value $I = 0.995$ was used for identifying a clear-sky day. The AERONET data are already selected for clear sky at the instance of measurement.

At this stage, all pre-processing operations on the two databases have been applied. Next the two datasets have to be merged. The merging is performed with the instantaneous values, i.e. no averaging is applied on any of the parameters. In order to combine the two datasets, only timestamps when data from both archives is available were retained. The resulting dataset inherits the time resolution from the AERONET data, i.e. 15-minute. The procedure for meshing the BSRN and AERONET datasets is described in more detail in [11]. The final dataset comprises data recorded under clear-sky conditions from overlapping time periods with records from both networks.

The algorithm guarantees that all selected days are sunny and quasi-cloudless during the whole day, leading to relatively few gaps in the measured data. For some stations the overlapping time intervals from the two archives are very brief, resulting in a small number of valid data. A lower threshold of 4000 data lines per station was imposed. The resulting dataset, comprising 12 locations spread around the world, is summarized in Table 1. The final dataset, after further selection, is described in Sec. 3.1.

TABLE 1. Source of data under which this study was conducted indicating the main Köppen-Geiger climate class and a summary-statistics (mean, first quartile Q1 and maximum value) of the Ångström turbidity parameter $\beta$ dataset at each station. $N$ is the number of data entries per station.

| No | Location (Country) | BSRN index | BSRN station | | | Climate | $N$ | $\beta_{Q1}$ | $\beta_{mean}$ | $\beta_{max}$ |
| | | | Lat. [deg] | Long. [deg] | Alt. [m] | | | | | |
| --- | --- | --- | --- | --- | --- | --- | --- | --- | --- | --- |
| 1 | Boulder (USA) | BOS | 40.125 | -105.237 | 1689 | BSk | 15506 | 0.011 | 0.023 | 0.310 |
| 2 | Cener (Spain) | CNR | 42.816 | -1.601 | 471 | Cfb | 8404 | 0.015 | 0.036 | 0.499 |
| 3 | Darwin (Australia) | DAR | -12.425 | 130.891 | 30 | Aw | 8141 | 0.035 | 0.054 | 0.214 |
| 4 | Fukuoka (Japan) | FUA | 33.582 | 130.376 | 3 | Cfa | 11649 | 0.062 | 0.133 | 0.887 |
| 5 | Gobabeb (Namibia) | GOB | -23.561 | 15.042 | 407 | Bwh | 77952 | 0.019 | 0.040 | 0.436 |
| 6 | Izana (Spain) | IZA | 28.309 | -16.499 | 2373 | Csb | 54441 | 0.006 | 0.043 | 1.065 |
| 7 | Langley Research Center (USA) | LRC | 37.104 | -76.387 | 3 | Cfa | 9314 | 0.015 | 0.030 | 0.462 |
| 8 | Palaiseau (France) | PAL | 48.713 | 2.208 | 156 | Cfb | 9476 | 0.027 | 0.054 | 0.514 |
| 9 | Sioux Falls (USA) | SFX | 43.730 | -96.620 | 473 | Dfa | 6153 | 0.022 | 0.036 | 0.223 |
| 10 | Tamanrasset (Algeria) | TAM | 22.790 | 5.529 | 1385 | Bwh | 16300 | 0.033 | 0.109 | 1.382 |
| 11 | Toravere (Estonia) | TOR | 58.254 | 26.462 | 70 | Dfb | 6606 | 0.028 | 0.055 | 0.839 |
| 12 | Xianghe (China) | XIA | 39.754 | 116.962 | 32 | Dwa | 15674 | 0.075 | 0.278 | 1.744 |

*2.2 Quantifying aerosol-induced losses in the solar potential*

In order to quantify the loss in the solar potential due to atmospheric aerosols, the measured solar irradiance in a given location must be compared with the corresponding value estimated under a reference scenario. Climatological values, which basically represent a long-time average of the given parameter, are commonly used as reference. Such a baseline is useful for showing biases or for quantifying what happens under extreme deviations from average conditions. For highlighting the loss in the overall solar potential, it is more appropriate to use some type of *ideal* scenario as a reference. The straightforward choice would be to consider an atmosphere without aerosols or a continental background value. However, neither of these options represent a realistic baseline for a specific location, especially for PV plants installed close to urban environments. This is explained next.

The contribution of aerosols to the extinction of a solar beam through the atmosphere is captured by the aerosol optical depth (AOD). In solar energy applications the AOD is further expressed in terms of the wavelength of the incoming beam through the Ångström equation:

$$AOD(\lambda) = \beta \lambda^{-\alpha}, \qquad (1)$$

where α and β are the Ångström coefficients, and the wavelength λ is expressed in μm. The Ångström Exponent α is inversely linked to the aerosol dynamic diameter, while the Ångström turbidity β, although not directly linked to any measurable quantity, captures information about particle number density.

Figure 1 shows a histogram – in log scale – of the Ångström turbidity parameter for the considered dataset. We observe that the distribution is bimodal, naturally separating into two quasi-lognormal distributions, which is characteristic for aerosols. Desert dust and urban industrial aerosols are captured in the higher mode. The smaller mode comprises, among others, the local aerosol background. It must be noted that the magnitude of measurement uncertainties is also of this order of magnitude [25].

Visual inspection of Fig.1 shows that the use of the climatological value (defined as the arithmetic mean of multi-year measurements: $\beta_{mean}$) would only capture a small part of the aerosol distribution, making it inadequate for quantifying the losses in the solar potential. A more physically based choice would be the minimum value between the two modes in the distribution in Fig. 1. For the whole dataset the minimum is at $\beta_{Q1} = 0.01$ (compare to $\beta_{mean} = 0.054$; see the dashed lines in Fig.1), which approximately lies at the first quartile of the data in terms of Ångström turbidity. Two impediments arise in choosing the exact location of the central minimum as a reference for each station: 1) for particular locations other small maxima appear, which make the definition of the minimum not straightforward, and 2) multi-year data is not always available to form a distribution which is statistically representative for each location. Hence a compromise was opted for here, with the directive of simplicity of implementation.

In this study, the first quartile (Q1) of the Ångström turbidity parameter – based on the measurements at each station – was chosen as representative of ideal conditions. This choice of threshold is partly subjective, used to illustrate the method. In a realistic study, the threshold should be set by analyzing the long-term aerosol trends and/or aerosol composition to separate out the various contributions, depending on the study goal. Figure 1 shows that this choice for the reference approximately partitions the two modes. In other words, we are assessing the loss in the solar potential that results from the presence of aerosols, as compared to the expected values that would occur in the presence of only the local aerosol background.

As the threshold is considered in terms of instantaneous values, and considering the strong aerosol seasonality, this setup is useful only for performing multi-year studies where the seasonal variations are averaged over. This condition is not strictly fulfilled at stations where long-term changes occur in the aerosol regime. For example, as explored in many studies, due to increasing pollution levels, parts of China have experienced a declining mean surface solar irradiance throughout the last decades [26], and showing an unprecedented increase after the implementation of the Clean Air Action in 2013 [27]. Indeed, this changing aerosol background can be simulated using the method introduced here. Nonetheless, as the goal of the present study is only to showcase the functionality of the proposed method, considerations about changing long-term aerosol trends and complex aerosol compositions are neglected here.

The main categories of aerosols represented in our dataset are considered as Urban-Industrial (UI) and Desert Dust (DD). Often an intermediate – 'mixed' – category is also considered (see e.g., [11], [28], [29] for a more detailed discussion). Categorization of aerosol type based on remotely sensed aerosol optical properties is, to some degree, a subjective task and can lead to unreliable results [30]. Nonetheless, this is the current practice in solar energy literature. Here, a very simple partitioning in terms of the Ångström exponent is used, namely DD: $\alpha < 0.8$ and UI: $\alpha > 0.8$ (see Fig. 1).

Further care should be taken when analyzing the impact of aerosols resulting from combustion and other burning processes. Values of $\alpha$ higher than 1.5, together with a low single scattering albedo, are often indicative of biomass burning products. On the other hand, the single scattering albedo of urban-industrial aerosols ranges from low (i.e., black smoke) to very high (i.e. white smoke). Because the single scattering albedo strongly influences the attenuation of the solar beam by aerosols, a further categorization of UI according to it should be made for precision. Furthermore the Ångström exponent is not able to distinguish well between aerosol types at low AOD. Nonetheless, for the purpose of this preliminary study we maintain only the simplistic categorization in terms of $\alpha$ into DD and UI classes, roughly separating aerosols into dust and combustion products. With this definition, the UI class is taken to include all aerosols resulting from burning processes, whether anthropogenic or biological. Marine aerosols are not a dominant aerosol type at any of the stations analyzed.

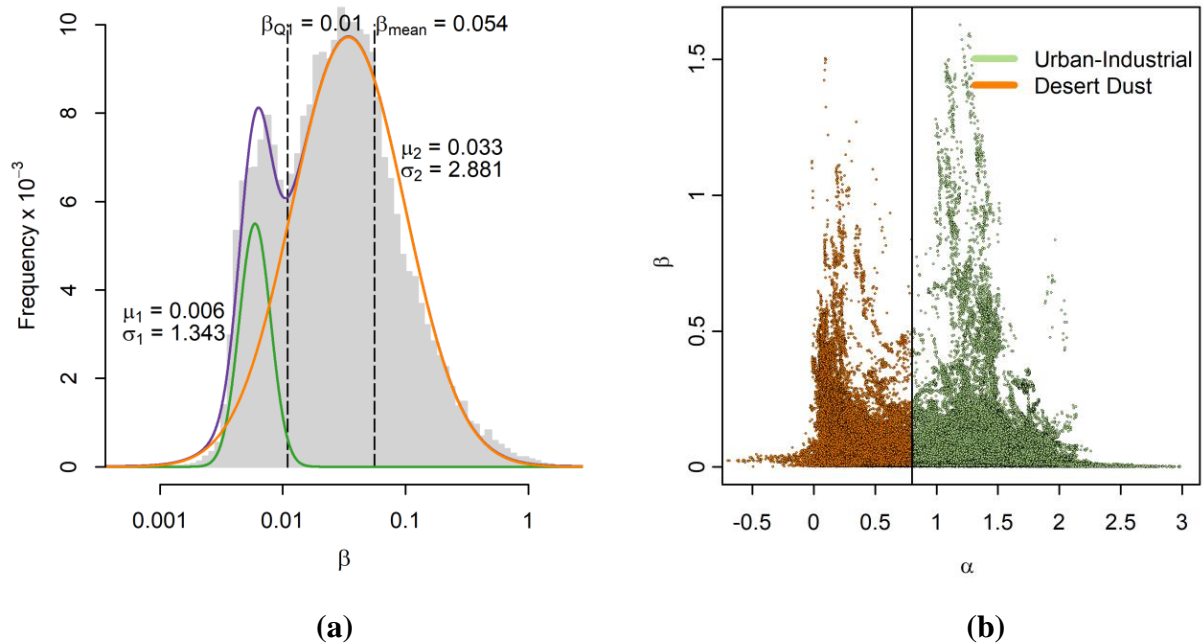

(a) (b)

FIGURE 1 (a) Histogram of Ångström turbidity parameter for the whole dataset. Overimposed on the data are lognormal curves for the two modes. $\mu$ and $\sigma$ represent the two parameters of the lognormal distribution, namely the median of the data and the geometrical standard deviation [31]. (b) Scatterplot of the Ångström turbidity parameter as a function of the Ångström exponent over the whole dataset. The classification in terms of major aerosol types is highlighted.

Assuming clear sky conditions, the loss in the daily solar potential can be defined as:

$$\Delta H = H_{m,ref} - H_m = \int_{t_{sr}}^{t_{ss}} G_{m,ref} dt - \int_{t_{sr}}^{t_{ss}} G_m dt, \quad (2)$$

where $G_m$ is the measured solar irradiance, while $G_{m,ref}$ represents the same quantity but under the reference conditions. $t_{sr}$ and $t_{ss}$ defines the sunrise and sunset times, respectively. Since $G_{m,ref}$ is not accessible experimentally, it must be modelled. While, $G_{ref}$ is the natural substitute for $G_{m,ref}$, this quantity is potentially strongly model-dependent. If the clear-sky model systematically under-/over-estimates the solar irradiance, this would lead to under-/over-estimation of the losses. A quantity that is less dependent on the clear-sky model is desirable.

In this study, we propose the following procedure for assessing the aerosol-induced loss in daily solar irradiance:

$$\begin{aligned}\Delta H &= \int_{t_{sr}}^{t_{ss}} G_{ref} k_{cs} dt - \int_{t_{sr}}^{t_{ss}} G_m dt \\ &= \int_{t_{sr}}^{t_{ss}} G_m \left( \frac{G_{ref}}{G_\beta} - 1 \right) dt,\end{aligned} \quad (3)$$

where $k_{cs} = G_m / G_\beta$ represents the clear-sky index. $G_\beta$ and $G_{ref}$ are the global horizontal solar irradiance estimated with a parametric clear-sky model taking as input the measured and reference values of the Ångström turbidity parameter $\beta$, respectively. $G_{ref} k_{cs}$ is proposed as an estimate for $G_{m,ref}$ in Eq. (2). The inclusion of the clear-sky index in the second term of Eq. (3) is proposed to adjust for possible model biases. It is well known that parametric models show systemic biases as a function of aerosol loading, especially significant at high turbidity [32]. By definition, this bias has the same sign at different values of the Ångström turbidity coefficient. If the model underestimates the measurements at $\beta_{Q1}$, the $k_{cs}$ (evaluated at $\beta_m$) is also above unity and counteracts the model bias by increasing the estimated irradiance. The situation is the inverse for model overestimation, always leading to a better estimate in the presence of the clear-sky index. If the solar irradiance model would be 100% accurate, then the clear-sky index would be unity ($k_{cs} = 1$), and the operation would be superfluous.

The fractional quantity in Eq. (3) is estimated with a clear-sky model as:

$$\frac{G_{ref}}{G_{\beta}} = \frac{G_{\text{model}}(h, l_i, \phi_j, \beta_{Q1})}{G_{\text{model}}(h, l_i, \phi_j, \beta_m)}, \quad \begin{matrix} l_i = \{l_w, l_{O3}, l_{NO2}, l_g, l_R, ...\} \\ \phi_j = \{\alpha, \tilde{\omega}, g, ...\} \end{matrix} \tag{4}$$

where $h$ is the Sun elevation angle, $l_i$ are the columnar values of the main atmospheric species besides aerosols (precipitable water vapor, ozone, nitrogen dioxide, mixed gases, Rayleigh scattering etc.), and $\phi_j$ are aerosol optical properties (Ångström exponent, single-scattering albedo, asymmetry factor etc.). Considering that all the input parameters for both $G_\beta$ and $G_{ref}$, besides the Ångström turbidity $\beta$, are the same measured values, Eq. (3) singles out the impact of aerosols. Furthermore, the Ångström exponent $\alpha$, single scattering albedo $\tilde{\omega}$, and all other physical quantities that determine the aerosol type are also fixed. Thus, although $\beta$ is not directly linkable to the particle number density (which is a measurable quantity), with the above assumptions we interpret variations in $\beta$ exclusively due to changes in aerosol particle number. Hence, with this interpretation, Eq. (3) quantifies aerosol-related solar irradiance losses that are due to changes in the atmospheric aerosol loading, while keeping the aerosol type constant.

Relative losses are computed as:

$$\Delta H_r = \frac{\Delta H}{H_{ref}} \times 100 \tag{5}$$

If losses in the direct-normal (or any other) solar potential are desired, these can be obtained by directly replacing the global horizontal irradiance with the desired quantity in Eqs. (2)-(5).

## 3. RESULTS AND ANALYSIS

### *3.1. Representative months*

A general view of the results across our dataset can be obtained by aggregating data by months, across all years at each station. Seasonal variations in the solar irradiance are clearly more significant than long-term changes over the considered 10-year period. Table 2 shows the number of days that pass the criteria described in Sec. 2.1 for each station. Only days having more than 20 measurements are included. With a 15-minute resolution, this threshold signifies at least 5-hour with measurements per day. Days with fewer data entries must contain at least some intervals with heavy cloud cover and are thus eliminated.

The number of valid days at each station is listed in Table 2, broken down by months. It can be seen that most stations achieve between 10-30 valid days for the study period. Most stations present a balanced distribution among the months, with only some stations, like DAR and TOR, having some months with a high number of valid days and other months that fail to achieve the 7-day threshold. The unusually high number of valid days at stations GOB and IZA, reflect the relatively low amount of cloud cover that these locations experience (for a further discussion see Sec. 3.5 and Table A1 from Appendix A).

A certain number of days is required for the results to be considered minimally representative for a given month at a given station over the 10-year period. If this number is taken to be too small, spurious effects might influence our interpretation of the results. If the number is too high, it will preclude the valorization of a large part of the dataset. Opting for a compromise, we have taken the threshold to be 7 days. The months that pass the threshold are highlighted in Table 2. Note that because the threshold number is small its value does have an influence on the results. Only months with a high number of valid days should be considered strongly representative for a given location.

TABLE 2. Number of valid days for each station, per each month, across the 10-year period. Months surpassing/failing the selection threshold (see Sec. 3.1) are highlighted in green/orange.

| Month | 1 | 2 | 3 | 4 | 5 | 6 | 7 | 8 | 9 | 10 | 11 | 12 |
|---|---|---|---|---|---|---|---|---|---|---|---|---|
| BOS | 13 | 20 | 21 | 7 | 11 | 26 | 11 | 10 | 30 | 31 | 28 | 22 |
| CNR | 11 | 6 | 12 | 5 | 4 | 25 | 26 | 34 | 20 | 20 | 7 | 12 |
| DAR | - | - | 3 | 5 | 17 | 62 | 45 | 51 | 11 | 5 | - | - |
| FUA | 5 | 6 | 18 | 18 | 15 | 1 | 9 | 12 | 4 | 25 | 14 | 4 |
| GOB | 28 | 38 | 42 | 61 | 94 | 78 | 67 | 57 | 45 | 41 | 67 | 63 |
| IZA | 117 | 82 | 113 | 109 | 166 | 232 | 226 | 195 | 139 | 71 | 57 | 97 |
| LRC | 20 | 14 | 18 | 16 | 10 | 3 | 9 | 7 | 4 | 24 | 34 | 16 |
| PAL | 2 | 9 | 27 | 22 | 10 | 12 | 14 | 14 | 22 | 21 | 5 | - |
| SFX | 6 | 16 | 18 | 15 | 12 | 13 | 16 | 10 | 12 | 17 | 16 | 4 |
| TAM | 6 | 18 | - | 7 | 7 | 6 | 24 | 5 | 19 | 22 | 4 | 29 |
| TOR | - | 2 | 21 | 27 | 25 | 9 | 6 | 5 | 4 | 9 | - | - |
| XIA | 42 | 13 | 24 | 26 | 14 | 6 | 12 | 7 | 11 | 23 | 13 | 14 |

### 3.2. Clear-sky model: REST2v5

The choice for the clear sky solar irradiance model used in this study is the REST2v5 model [33], a model that uses two broad bands. REST2v5 is widely regarded as one of the most reliable clear-sky models available in the literature [34, 35]. Ground data from AERONET, described in Sec.

2.1. are used for all inputs of REST2v5: ozone column content $l_{O3}$, water vapor column content $w$, Ångström turbidity parameter $\beta$, Ångström exponent $\alpha$ and single scattering albedo $SSA$. The $SSA$ is obtained through the almucantar method and is recorded only a few times each day. Daily average $SSA$, based on the existing measurements, is considered here. No averaging is applied on any of the other parameters. The ground albedo $\rho$ is also taken from AERONET.

The reader is reminded that the AERONET data has a temporal resolution of 15-minute, and it is occasionally sparse. In order to evaluate daily solar irradiance values, the model input variables and $k_{cs}$ were interpolated onto a finer temporal grid. Figure 2 illustrates this simple linear interpolation. At the ends of the interval, the closest measured value is used for each variable up to sunrise/sunset.

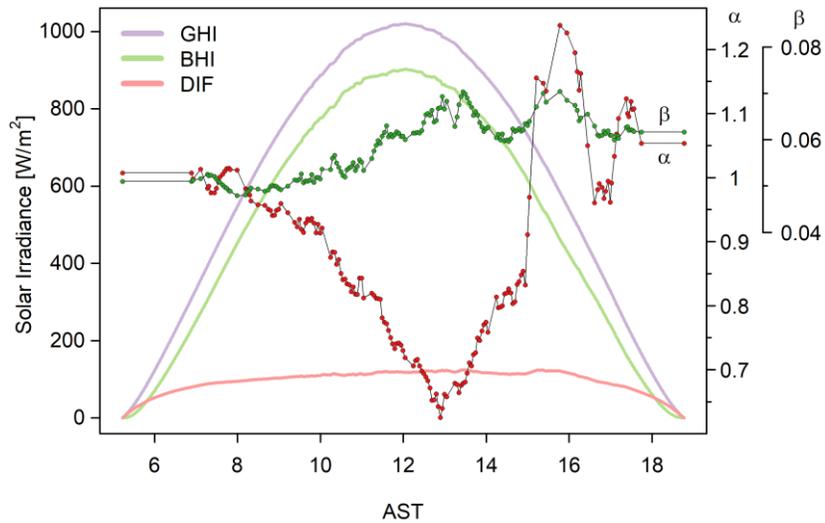

FIGURE 2. Measured (dots) and interpolated (solid black lines) values of the Ångström turbidity coefficient $\beta$ and the Ångström exponent $\alpha$ with respect to the apparent solar time (AST), illustrated for a generic clear sky day in the dataset (LRC, 2018-05-01). The estimated values of the three horizontal solar irradiance components – global (GHI), beam (BHI) and diffuse (DIF) – are also shown.

### 3.3. Average daily losses

The solar irradiation values obtained with REST2v5, as described in Sec. 2.2 and 3.2, estimate the solar irradiance in the hypothesis of permanently clear sky. In other words, the daily and monthly values that are computed based on these estimates represent the amount of solar energy that would be measured at the ground if the whole period would be clear of clouds. The estimated daily losses in the solar potential as described through Eq. (3), i.e the estimated reference minus the measured global horizontal solar irradiation (GHI) values, are listed in Table. 3.

TABLE 3. Differences $\Delta H$ [Wh/m$^2$] (Eq. 3) between daily GHI estimated with REST2v5 using the measured and the baseline values for the Ångström turbidity parameter. Relative losses $\Delta H_r$ [%] (Eq. 5), calculated with respect to the reference daily GHI, are also listed. Cells are shaded from light to dark according to percentage intervals: below 0%; [0%, 2.5%); [2.5%, 5%); [5%,10%); [10%, 25%); above 25%. Missing values are designated as NA.

| Month | | 1 | 2 | 3 | 4 | 5 | 6 | 7 | 8 | 9 | 10 | 11 | 12 |
|---|---|---|---|---|---|---|---|---|---|---|---|---|---|
| BOS | $\Delta H$ | -7.3 | 9.53 | 45.71 | 89.51 | 59.25 | 97.25 | 81.13 | 29.29 | 76.82 | 20.52 | -2.53 | -0.83 |
| | $\Delta H_r$ | -0.24 | 0.21 | 0.75 | 1.18 | 0.67 | 1.09 | 0.97 | 0.4 | 1.26 | 0.43 | -0.07 | -0.03 |
| CNR | $\Delta H$ | 12.78 | NA | 67.16 | NA | NA | 132.53 | 93.52 | 67.48 | 61.61 | -16.69 | 1.82 | -0.77 |
| | $\Delta H_r$ | 0.5 | NA | 1.3 | NA | NA | 1.56 | 1.13 | 0.91 | 1.09 | -0.42 | 0.07 | -0.03 |
| DAR | $\Delta H$ | NA | NA | -66.43 | -31.19 | 20.62 | 72.86 | 64.22 | NA | NA | NA | NA | NA |
| | $\Delta H_r$ | NA | NA | -0.97 | -0.48 | 0.35 | 1.3 | 1.13 | NA | NA | NA | NA | NA |
| FUA | $\Delta H$ | NA | NA | 452.87 | 279.06 | 235.2 | NA | 156.34 | 239.81 | NA | 34.6 | 14.44 | NA |
| | $\Delta H_r$ | NA | NA | 7.12 | 3.72 | 3.02 | NA | 2.01 | 3.24 | NA | 0.69 | 0.36 | NA |
| GOB | $\Delta H$ | 113.71 | 78.97 | 19.3 | 14.57 | 14.85 | 7.7 | 32.99 | 124.58 | 175.03 | 139.5 | 73.77 | 53.72 |
| | $\Delta H_r$ | 1.26 | 0.92 | 0.26 | 0.24 | 0.29 | 0.16 | 0.67 | 2.14 | 2.45 | 1.74 | 0.82 | 0.58 |
| IZA | $\Delta H$ | 29.43 | 4.85 | 21.56 | 43.31 | 28.77 | 78.94 | 197.44 | 159.38 | 29.74 | 9.76 | 4.71 | 7.43 |
| | $\Delta H_r$ | 0.58 | 0.08 | 0.29 | 0.5 | 0.31 | 0.83 | 2.12 | 1.83 | 0.38 | 0.15 | 0.09 | 0.16 |
| LRC | $\Delta H$ | -2.04 | 17.49 | 46.11 | 54.48 | 138.63 | NA | 110.64 | 119.87 | NA | 33.44 | 5.64 | 13.25 |
| | $\Delta H_r$ | -0.06 | 0.38 | 0.75 | 0.72 | 1.67 | NA | 1.54 | 1.62 | NA | 0.7 | 0.16 | 0.46 |
| PAL | $\Delta H$ | NA | 70.81 | 165.17 | 168.46 | 76.53 | 35.5 | 107.46 | 78.4 | 18.87 | 65.02 | NA | NA |
| | $\Delta H_r$ | NA | 2.33 | 3.67 | 2.57 | 0.96 | 0.42 | 1.3 | 1.14 | 0.36 | 1.73 | NA | NA |
| SXF | $\Delta H$ | NA | -6.04 | 21.4 | 63.86 | 60.8 | 46.23 | 89.35 | 52.32 | 89.55 | 3.77 | -3.15 | NA |
| | $\Delta H_r$ | NA | -0.15 | 0.4 | 0.89 | 0.73 | 0.53 | 1.09 | 0.79 | 1.55 | 0.09 | -0.11 | NA |
| TAM | $\Delta H$ | NA | 114.05 | 316.5 | 291.76 | NA | 629.67 | NA | 630.53 | 341.32 | NA | 138.79 | NA |
| | $\Delta H_r$ | NA | 2.08 | 5.46 | 4.83 | NA | 10.17 | NA | 8.64 | 5.78 | NA | 2.86 | NA |
| TOR | $\Delta H$ | NA | -14.18 | 11.33 | 88.91 | 29.85 | NA | NA | NA | 162.18 | NA | NA | NA |
| | $\Delta H_r$ | NA | -0.67 | 0.32 | 1.54 | 0.39 | NA | NA | NA | 4.28 | NA | NA | NA |
| XIA | $\Delta H$ | 136.69 | 646.33 | 447.38 | 850.6 | 586.84 | NA | 261.26 | 115.29 | 191.08 | 515.74 | 188.84 | 253.44 |
| | $\Delta H_r$ | 4.71 | 17.44 | 8.35 | 11.71 | 7.41 | NA | 3.21 | 1.64 | 3.36 | 11.98 | 6.74 | 12.42 |

TABLE 4. Differences ΔH [Wh/m²] (Eq. 3) between daily DNI estimated with REST2v.5 using the measured and the reference values for the Ångström turbidity parameter. Relative losses [ΔH$_r$(%)] (Eq. 5), calculated with respect to the baseline daily irradiation, are also listed. Cells are shaded from light to dark according to percentage intervals: below 0%; [0%, 2.5%); [2.5%, 5%); [5%,10%); [10%, 25%); above 25%. Missing values are designated as NA.

| Month | | 1 | 2 | 3 | 4 | 5 | 6 | 7 | 8 | 9 | 10 | 11 | 12 |
|---|---|---|---|---|---|---|---|---|---|---|---|---|---|
| BOS | ΔH | -70.09 | 72.39 | 513.46 | 947.07 | 585 | 959.81 | 763.97 | 1041.05 | 756.64 | 177.18 | -50.69 | -15.96 |
|  | ΔH$_r$ | -0.89 | 0.77 | 4.76 | 8.12 | 4.74 | 7.73 | 6.53 | 9.54 | 7.39 | 1.88 | -0.6 | -0.21 |
| CNR | ΔH | 137.6 | NA | 700.32 | NA | NA | 1028.48 | 695.52 | 610.01 | 454.78 | 434.27 | -74.19 | -146.36 |
|  | ΔH$_r$ | 2.09 | NA | 7.64 | NA | NA | 8.75 | 6.08 | 5.63 | 4.81 | 5.38 | -1.13 | -2.4 |
| DAR | ΔH | NA | NA | -347.01 | -276.82 | 129.13 | 418.92 | 399.33 | NA | NA | NA | NA | NA |
|  | ΔH$_r$ | NA | NA | -3.87 | -3.14 | 1.54 | 5.16 | 4.94 | NA | NA | NA | NA | NA |
| FUA | ΔH | NA | NA | 2332.55 | 1592.85 | 1442.04 | NA | 855.62 | 1315.94 | NA | 215.41 | 31.73 | NA |
|  | ΔH$_r$ | NA | NA | 26.8 | 17.15 | 15 | NA | 9.6 | 15.26 | NA | 2.93 | 0.48 | NA |
| GOB | ΔH | 1037.72 | 742.43 | 289.41 | 155.5 | 183.93 | 72.74 | 273.21 | 921.29 | 1120.9 | 1236.8 | 868.04 | 712.13 |
|  | ΔH$_r$ | 9.04 | 6.73 | 2.91 | 1.73 | 2.23 | 0.91 | 3.32 | 10.45 | 11.45 | 11.9 | 7.54 | 6.07 |
| IZA | ΔH | 184.25 | 82.75 | 306.46 | 490.4 | 407.96 | 727.96 | 1780.81 | 1528.96 | 368.28 | 93.79 | 66.27 | 75.43 |
|  | ΔH$_r$ | 1.8 | 0.76 | 2.59 | 3.88 | 3.1 | 5.5 | 13.68 | 12.38 | 3.15 | 0.85 | 0.65 | 0.77 |
| LRC | ΔH | -15.97 | 231.86 | 515.9 | 626.22 | 1380.53 | NA | 1042.71 | 1042.82 | NA | 404.95 | 62.87 | 166.1 |
|  | ΔH$_r$ | -0.21 | 2.65 | 5.13 | 5.68 | 12.14 | NA | 9.78 | 10.03 | NA | 4.68 | 0.82 | 2.4 |
| PAL | ΔH | NA | 707.86 | 1354.55 | 1325.2 | 774.13 | 422.39 | 764.72 | 724.72 | 205.84 | 525.96 | NA | NA |
|  | ΔH$_r$ | NA | 10.12 | 16.13 | 12.79 | 6.78 | 3.58 | 6.61 | 7.02 | 2.28 | 6.86 | NA | NA |
| SXF | ΔH | NA | -120.38 | 261.24 | 694.71 | 698.38 | 554.52 | 671.99 | 733.31 | 762.36 | 52.48 | -46.99 | NA |
|  | ΔH$_r$ | NA | -1.49 | 2.83 | 6.41 | 5.98 | 4.68 | 6.03 | 7.46 | 8.18 | 0.65 | -0.69 | NA |
| TAM | ΔH | NA | 710.5 | 2110.83 | 2111.67 | NA | 4074.72 | NA | 3512.5 | 2442.43 | NA | 949.71 | NA |
|  | ΔH$_r$ | NA | 7.57 | 21.77 | 21.01 | NA | 37.73 | NA | 32.71 | 24.78 | NA | 11.13 | NA |
| TOR | ΔH | NA | -82.3 | -110.6 | 762.68 | 292.05 | NA | NA | NA | 1284.8 | NA | NA | NA |
|  | ΔH$_r$ | NA | -1.42 | -1.43 | 7.33 | 2.4 | NA | NA | NA | 15.47 | NA | NA | NA |
| XIA | ΔH | 622.09 | 2535.87 | 1637.57 | 3350.36 | 2571.47 | NA | 795.17 | 576.11 | 897.25 | 1911.76 | 867.23 | 1418.84 |
|  | ΔH$_r$ | 11.01 | 39 | 20.73 | 35.25 | 26.11 | NA | 8.37 | 6.56 | 11.36 | 27.75 | 16.07 | 32.01 |

As a general overview, in most cases the daily losses in terms of GHI are below 2-3%, although in particular months they can be much higher. There are a few stations which present persistently highly losses due to aerosols, namely Tamanrasset, Algeria (TAM), Xianghe, China (XIA), and to a lesser degree Fukuoka, Japan (FUA). The largest daily loss is observed at XIA during April, at 850.6 Wh/m$^2$, while the highest relative loss is seen at the same station during February, at 17.4%. Losses above 10% are also observed in other months at XIA (Apr, Oct, and Dec) and at TAM (Jun).

For concentrating solar power systems, the direct normal irradiance (DNI) is the relevant solar irradiance quantity. The losses in terms of DNI are listed in Table 4. We observe that they are, in general, notably higher than the GHI losses, with some stations experiencing losses higher than 25%. The highest losses are observed at TAM during June (4074.7 Wh/m$^2$), while in relative values they are at XIA during February (39%). TAM and XIA are the locations presenting the highest losses overall, however, even at FUA they can be as high as one quarter compared to the reference (26.8%, during March). Other months with losses higher than 20% of the potential are present at TAM (Mar, Apr, Aug, and Sep) and XIA (Mar, Apr, May, Oct, and Dec). The complex aerosol mix with high atmospheric loading at XIA is particularly interesting, being well studied in the literature [36].

The differences between the losses in terms of GHI and DNI are considerable. We can understand this by noting that part of the solar beam that is lost from DNI is scattered downward and still contributes to GHI. Thus, aerosols that have a high single-scattering albedo (like desert dust) show smaller relative losses in GHI than in DNI, while aerosols with a higher absorption ratio (like urban-industrial aerosols) present a higher loss in both GHI and DNI. It is important, thus, to look at the impact of aerosols as a function of aerosol type.

*3.4. Influence of aerosol type*

There is a stark contrast between the situation at the two mentioned stations, TAM and XIA. TAM station is located on the northern edge of the Sahara Desert in Algeria, the dominant aerosol being desert dust and the losses reflect the seasonality of dust storms. On the other hand, the XIA station is located close to the urban centers of Beijing and Tianjin in China, with the dominant aerosol being urban-industrial haze, the production of which is controlled by anthropogenic activity [37].

The AOD as a function of Ångström exponent for our data from these stations is shown in Figure 3. The dominant type of aerosol is clearly identifiable at each station, considering the classification from Sec. 2.2. (see also Fig. 1).

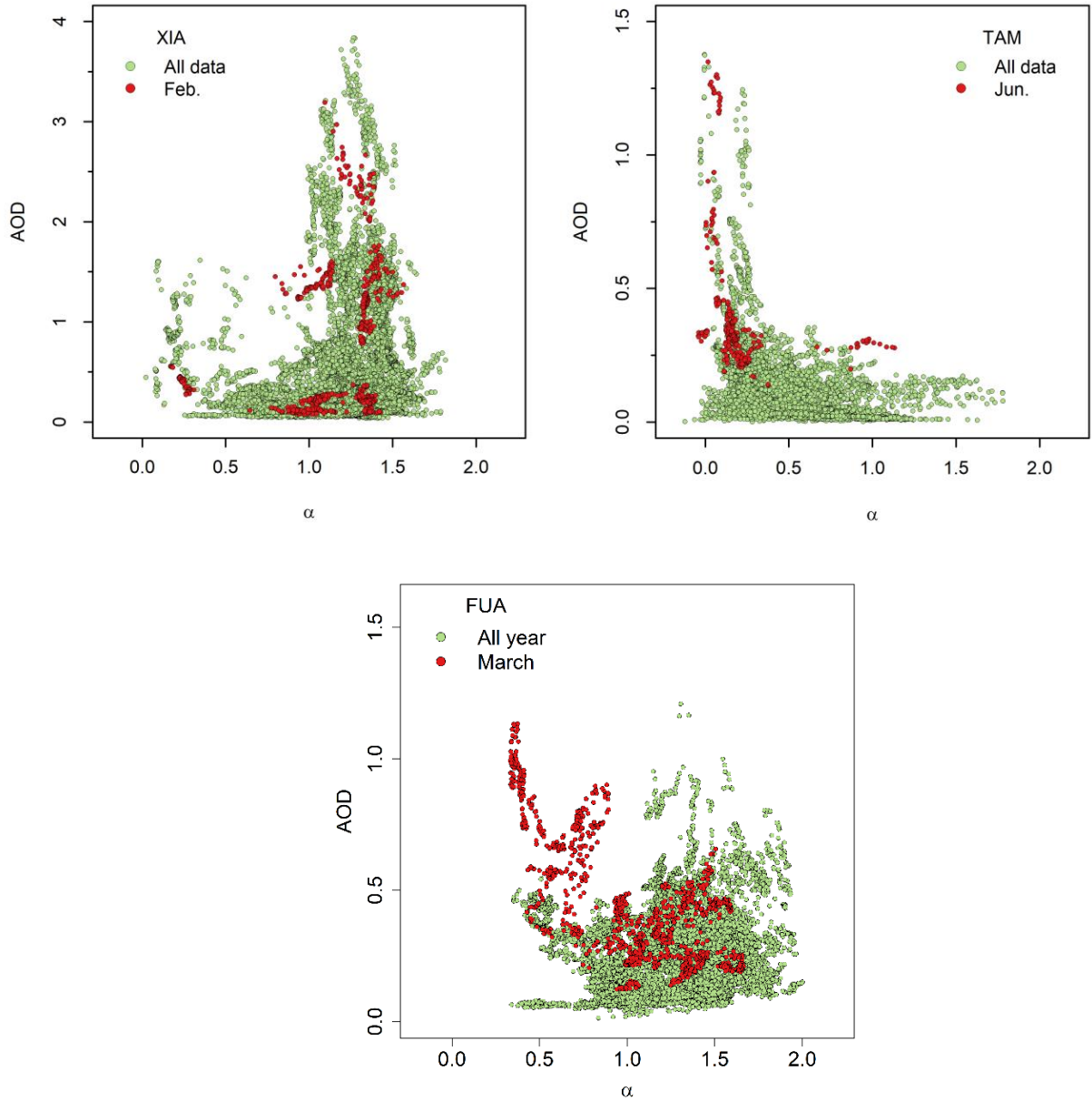

FIGURE 3. AOD as a function of Ångström exponent for the stations Tamanrasset (TAM), Xianghe (XIA), and Fukuoka (FUA) from our dataset. The data from the months with the highest daily losses in the solar potential are highlighted for each station (see. Tables 3 and 4). Note that with the classification from Sec. 2.2, $\alpha = 0.8$ is the threshold between DD and UI aerosols.

In the case of desert dust, there is no real way to prevent such losses. It is worth keeping in mind that worst case scenario climate change will come with widespread desertification [38], and, thus, such scenarios might be more common in the future. On the other hand, for locations dominated by urban-industrial aerosols, the losses are entirely manmade. The very high levels of pollution common around urban centers in quickly developing regions around the world—like parts of China—can be seen to have a very large impact on the solar potential, with losses in the DNI potential higher than 20% being present during several months at XIA (see Table 4).

There is an interesting midway situation for the station FUA. The city of Fukuoka, located in southwestern Japan, is part of a highly industrialized area, which is visible in the high levels of urban-industrial pollution shown in Fig. 3. At the same time, Japan is periodically affected by plumes of dust from the Gobi Desert. This is the case, in particular, for the month of March which is the month with the highest average aerosol presence. We see from the highlighted data points in Fig. 3 that the high AOD during March at FUA is actually generated as a combined effect of desert dust and high urban-industrial pollution.

It is instructive to compare the situation from August at TAM and May at XIA. In terms of GHI, the losses at TAM [630.5 Wh/m$^2$ (8.6%)] and XIA [586.8 Wh/m$^2$ (7.4%)] are similar, while in terms of DNI there is an almost 1000 Wh/m$^2$ difference between the two cases [TAM: 3512.5 Wh/m$^2$ (32.7%) vs. XIA: 2571.5 Wh/m$^2$ (26.1%)]. As noted in the previous section, this difference is due to the differing single scattering albedo of the aerosols predominant at the two stations. A more complex analysis of radiative impact by aerosol type could be obtained by using a more detailed aerosol categorization scheme, a task that is left for future work (see e.g. Ref. [39] for such an analysis for the XIA station).

### 3.5. Monthly and yearly average losses

Naturally, considering that our estimates refer only to clear-sky conditions, there will be very few months when they represent estimates for the real measured irradiances. In reality, a considerable number of days will be partially cloudy or overcast during most months.

Accounting for the simultaneous impact of clouds and aerosols is notoriously difficult because the effects of clouds and aerosols stack in non-linear ways. Aerosols act as cloud condensation nuclei, facilitating cloud formation and, thus, directly affecting cloud coverage [31]. The presence of aerosols can also affect other cloud properties, like the albedo and lifetime [40]. On the other hand, clouds impact the properties of aerosols through hygroscopic growth enhanced by the increase of water vapor brought by clouds [41].

Nonetheless, a very coarse estimation can be made by combining our approach with an Ångström equation to account for cloud coverage. An Ångström-type equation relates the measured solar irradiation with the sunshine duration σ under a given time interval, through a correlation of the form $H/H_{cs} = f(\sigma)$. In this study, the empirical equation developed in [42] was employed, namely:

$$H = H_{cs}(0.368 + 0.623\sigma). \tag{6}$$

σ is estimated based on measured DNI from the BSRN archive – the full set of measurements for the given period, not the final dataset which is screened for clear sky – using the following relation:

$$\sigma(\Delta T) = \int_{\Delta T} \xi(t) dt \tag{7}$$

$$\xi(t) = \begin{cases} 1 & \text{IF} \quad DNI > 120 \text{ W/m}^2 \\ 0 & \text{OTHERWISE} \end{cases} \tag{8}$$

The binary quantity $\xi(t)$ is called the sunshine number and tracks whether the Sun disk is covered by clouds or not at time $t$ [43, 44]. Eq. (8) defines $\xi$ according to the WMO sunshine criterion. The computed monthly averages of daily relative sunshine are listed in Table A1 from Appendix A.

Estimates of the monthly irradiation loss at each station are obtained through the daily losses computed in Sec. 3.3. combined with Eq. (6), as follows:

$$H_M^{cl+ae} = H_M^{ref} f(\sigma)(1 - \Delta H_r) = H_M^{cl}(1 - \Delta H_r), \tag{9a}$$

$$\Delta H_M^{ae} = H_M^{cl} - H_M^{cl+ae} = H_M^{ref} f(\sigma) \Delta H_r. \tag{9b}$$

The labels *cl* and *ae* signify clouds and aerosols. Hence, the notation $H_M^{cl+ae}$ means the monthly irradiation estimated in the presence of both clouds and aerosols, while $\Delta H_M^{ae}$ means the monthly loss in solar potential due exclusively to aerosols. Note that Eqs. (9) assume independence between the cloud and aerosols extinction processes on solar radiation, which is not strictly the case in reality as discussed above. Nonetheless, for monthly and yearly timescales it is considered a reasonable approximation for the purposes of this study.

The results are illustrated in Figure 4 for the stations TAM and XIA, the two stations from our dataset with the highest average turbidity. The monthly losses due to clouds are, naturally, always the largest. However, the aerosol-induced losses are not negligible and can be the same order of magnitude as that due to clouds. For example, during the months of Mar and Jun at TAM, the monthly aerosol losses are 8.9 kWh/m$^2$ and 14.4 kWh/m$^2$, compared to 16.5 kWh/m$^2$ and 44.3 kWh/m$^2$ due to clouds.

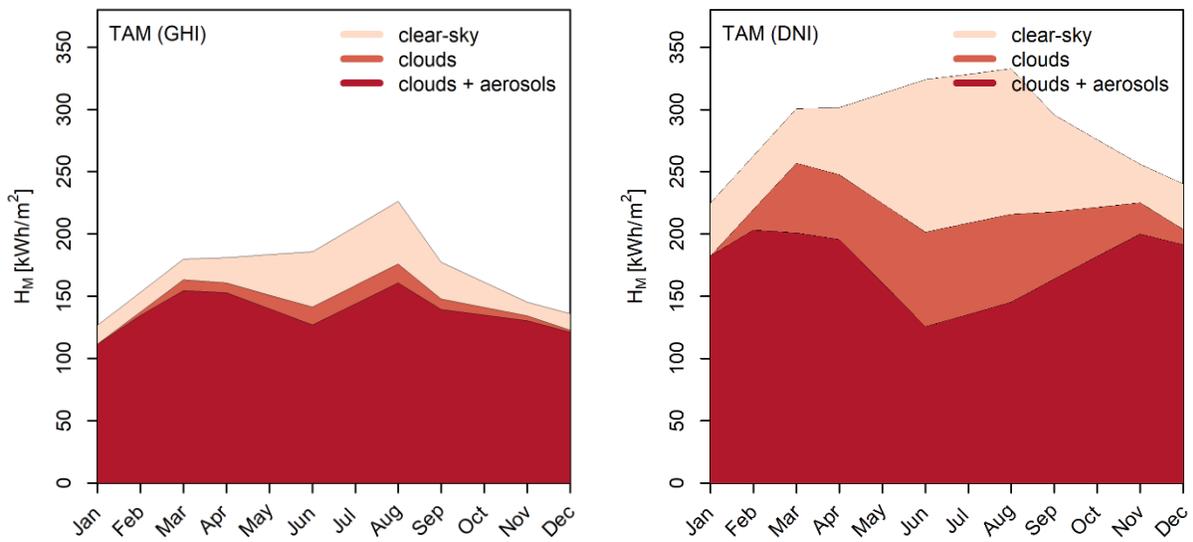

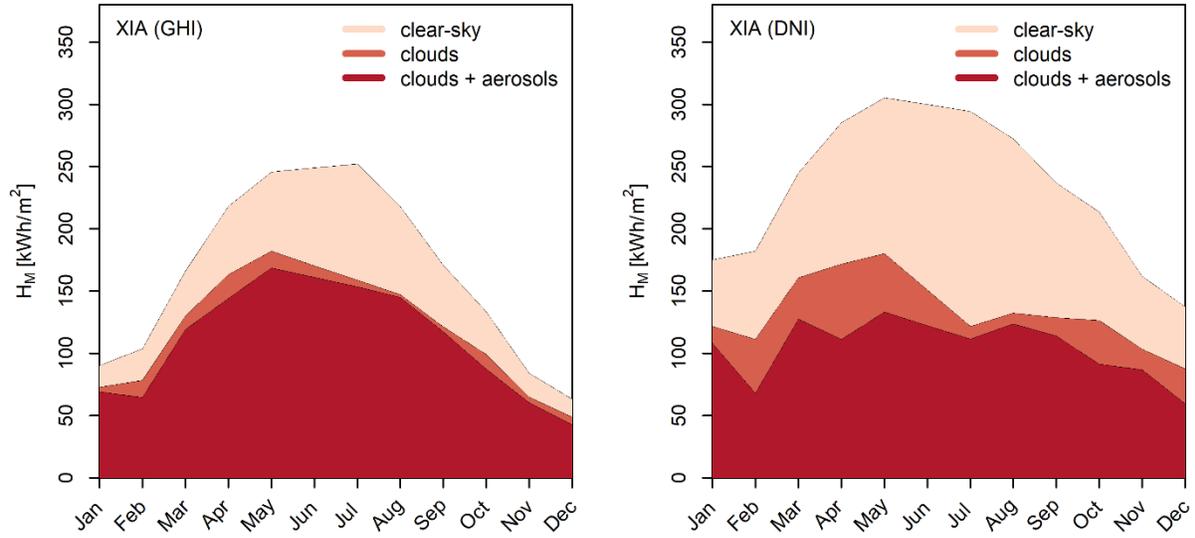

FIGURE 4. Estimates of average monthly global horizontal and direct normal solar irradiation for the stations TAM and XIA from our dataset, i) under reference clear-sky conditions $H_M^{ref}$, ii) in the presence of clouds $H_M^{cl}$, calculated with Eq. (6), and iii) in the presence of both clouds and aerosols $H_M^{cl+ae}$, estimated with Eq. (9a).

The monthly losses for all locations are listed in Table A2. The solar potential in the presence of clouds and the aerosol-induced losses are shown. In order to maximize the utility of the results, simple linear interpolation (with centered, left, and right differences, in this order) was used to estimate missing values. Interpolation was applied twice. Thus, most of the missing months are filled in, albeit with reduced confidence in the results. Interpolated values are italicized in Table A2. Only stations DAR and TOR contain missing values after interpolation. For the other stations, yearly totals were also included. The higher solar potential for the stations GOB and IZA is explained through the higher relative sunshine achieved at these stations, as shown in Table A1.

The yearly aerosol-induced losses, expressed as a percentage of the solar potential in the presence of clouds, is shown in Fig. 5. The highest losses are registered at FUA, TAM and XIA, amounting to 3.8%, 5.2%, and 7.2% of the yearly solar potential at each location, while at the other stations it ranges between 0.6-1.2%. Furthermore, the linear interpolation used for missing monthly values captures only the seasonal aerosol trend, missing episodes of particularly high atmospheric loading. This is potentially expected to lead to underestimation of the actual aerosol losses. In conclusion, the aerosol-induced reductions in the solar potential are considerable at many locations, and their omission or underestimation during the feasibility studies of PV projects can cause a substantial problem.

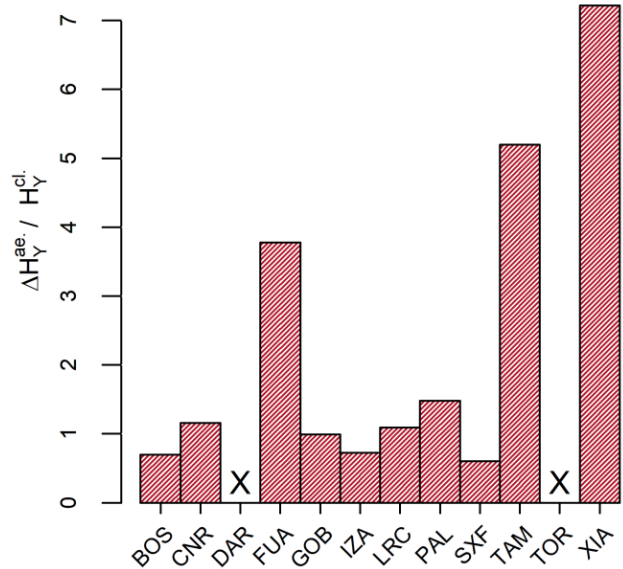

FIGURE 5. Yearly aerosol-induced reduction in the solar potential at the 12 stations from Table 1. Losses are expressed as a percentage of solar potential in the presence of clouds. Locations for which yearly values cannot be computed, due to missing monthly values, are marked with an X.

3.6 Discussion

The reduction in the solar potential due to aerosols is not a directly measurable quantity in realistic conditions, in the outside environment. Hence, the quality of the results can only be gauged by controlling the errors introduced by the different elements in the estimation procedure. In this case, the accuracy of the results is controlled by two factors, namely the accuracy of the data used and the accuracy of the estimation model. Furthermore, there is the issue of consistency between the choice of reference scenario and type of losses one is assessing. In this study, we have employed ground-based data which are the most reliable form of data at the present (usually considered 'ground truth') [45]. The physics-based REST2 clear-sky model was chosen to estimate the solar irradiance, considering its high performance established in the literature [46, 47]. Nonetheless, even REST2 sometimes presents artefacts in the estimation of DNI, as shown in Ref. [48]. Hence, some uncertainty is introduced by the use of a clear-sky model. The reference scenario was chosen to single out the impact of dust or combustion product aerosols above the local background. The ad-hoc way of determining the reference scenario as the Q1 of the locally measured aerosol data, namely Ångström turbidity, is a clear limitation of the method. For locations where detailed and long-term aerosol data is available, a more physically-based reference can be chosen based on an analysis of the data. For locations where such data is not available, the Q1 is a reasonable compromise.

The aerosol losses estimated at the 12 sites studied are consistent with other findings from the literature. Ref. [9], for example, found a 1.5kWh/m$^2$ loss in the solar potential due to aerosols over the most polluted regions of China. This finding is consistent with the losses reported through our method, where the losses range from 0.5 to 3.3kWh/m$^2$ from lowest to highest month, with a 1.56kWh/m2 average loss across a whole year.

The natural next step for this research is the application of the method developed here for satellite data to generate worldwide maps of aerosol losses for different scenarios. For this, the satellite product must first be calibrated in a way that is tailored for our method, based on ground data from AERONET or other sources. Another line of research will be the estimation of the contribution of different sources from the aerosol mix to the columnar aerosol extinction. For example, the study of the radiative impact of single extreme events, like a dust storm from the Sahara blanketing Southern Europe, similarly to the study in Ref. [49], will be possible.

4. CONCLUSION

In this paper, a procedure for estimating the reduction in the solar potential due to aerosols on different timescales was presented. Departing from the commonly threaded path of employing satellite data, here we use ground-measurements from two monitoring networks. An algorithm for combining data from the two repositories, AERONET and BSRN, which contain different types of data, with different temporal resolution, is described. Losses are assessed at 12 stations distributed across 5 continents.

Aiming to ensure a high degree of generality, the study was focused on the aerosol impact on collectable solar energy. Thus, the results are independent of solar plants' characteristics. A new methodology for estimating the average daily, monthly, and yearly losses in the solar potential due to aerosols was proposed. For highlighting the loss in the overall solar potential, a new *ideal* scenario was defined as a reference for the atmospheric aerosol background. The scenario is based on the first quartile of the Ångström turbidity coefficient dataset recorded at each station. With this procedure only one year of locally measured data is necessary, and the reference in easy to implement. The loss was evaluated from the measured solar potential in real atmospheric conditions, as compared to the expected values that would occur in the presence of only the local aerosol background. The specific impact of aerosol type (dust versus combustion aerosol) was studied.

The highest losses are registered where desert dust and/or urban industrial aerosols are abundant in the atmosphere: Fukuoka (Japan), Tamanrasset (Algeria), and Xianghe (China), with losses amounting to 3.8% (57.4 kWh/m$^2$/yr), 5.2% (90.7 kWh/m$^2$/yr), and 7.2% (103.7 kWh/m$^2$/yr) of the solar potential. These effects are very high and could translate into huge costs if omitted or underestimated during the planning phase of PV projects. The largest currently operating solar power plants in the world surpass 1GW nominal capacity and cover tens of square kilometers. For such a plant, even a 1% underestimation in the collectable solar energy translates into massive energy and financial losses.

APPENDIX A

Table A1. Monthly averages of daily relative sunshine for the stations listed in Table 1.

|     | 1 | 2 | 3 | 4 | 5 | 6 | 7 | 8 | 9 | 10 | 11 | 12 |
|-----|---|---|---|---|---|---|---|---|---|----|----|----|
| BOS | 0.675 | 0.667 | 0.661 | 0.631 | 0.601 | 0.750 | 0.727 | 0.711 | 0.721 | 0.695 | 0.686 | 0.653 |
| CNR | 0.343 | 0.315 | 0.408 | 0.551 | 0.527 | 0.642 | 0.735 | 0.757 | 0.673 | 0.572 | 0.347 | 0.403 |
| DAR | 0.437 | 0.492 | 0.530 | 0.712 | 0.790 | 0.937 | 0.913 | 0.944 | 0.873 | 0.777 | 0.640 | 0.523 |
| FUA | 0.374 | 0.388 | 0.510 | 0.525 | 0.541 | 0.339 | 0.460 | 0.541 | 0.485 | 0.522 | 0.456 | 0.342 |
| GOB | 0.830 | 0.792 | 0.863 | 0.903 | 0.964 | 0.952 | 0.939 | 0.895 | 0.850 | 0.850 | 0.872 | 0.850 |
| IZA | 0.857 | 0.840 | 0.834 | 0.848 | 0.940 | 0.970 | 0.969 | 0.925 | 0.871 | 0.741 | 0.750 | 0.797 |
| LRC | 0.500 | 0.465 | 0.523 | 0.625 | 0.600 | 0.648 | 0.688 | 0.661 | 0.543 | 0.613 | 0.580 | 0.475 |
| PAL | 0.241 | 0.306 | 0.386 | 0.510 | 0.471 | 0.502 | 0.548 | 0.526 | 0.535 | 0.408 | 0.258 | 0.290 |
| SXF | 0.510 | 0.583 | 0.538 | 0.585 | 0.578 | 0.673 | 0.782 | 0.707 | 0.695 | 0.629 | 0.612 | 0.465 |
| TAM | 0.898 | 0.836 | 0.855 | 0.821 | 0.717 | 0.622 | 0.668 | 0.648 | 0.737 | 0.835 | 0.879 | 0.880 |
| TOR | 0.169 | 0.261 | 0.373 | 0.494 | 0.574 | 0.514 | 0.541 | 0.505 | 0.390 | 0.222 | 0.112 | 0.143 |
| XIA | 0.694 | 0.612 | 0.656 | 0.602 | 0.590 | 0.479 | 0.413 | 0.486 | 0.543 | 0.593 | 0.639 | 0.637 |

Table A2. Estimated monthly global horizontal irradiance in the presence of clouds $H_M^{cl}$ and monthly losses in solar potential due exclusively to aerosols $\Delta H_M^{ae}$ in each location. Missing values were interpolated linearly (italicized).

|     |     | Monthly | | | | | | | | | | | | Yearly |
|-----|-----|---|---|---|---|---|---|---|---|---|----|----|----|--------|
|     | Month | 1 | 2 | 3 | 4 | 5 | 6 | 7 | 8 | 9 | 10 | 11 | 12 | |
| BOS | $H_M^{cl}$ | 76.5 | 100.6 | 149.2 | 175.1 | 205.2 | 225.9 | 214.1 | 184.2 | 150.2 | 119 | 82.4 | 67.1 | 1749.6 |
|     | $\Delta H_M^{ae}$ | -0.2 | 0.2 | 1.1 | 2.1 | 1.4 | 2.5 | 2.1 | 0.7 | 1.9 | 0.5 | -0.1 | 0 | 12.2 |
| CNR | $H_M^{cl}$ | 46.3 | *73.3* | 100.3 | *140.4* | *180.6* | 197.3 | 214.1 | 194 | 134.2 | 90.1 | 45.7 | 43.5 | 1459.8 |
|     | $\Delta H_M^{ae}$ | 0.2 | *0.8* | 1.3 | *2.5* | *3.7* | 3.1 | 2.4 | 1.8 | 1.5 | -0.4 | 0 | 0 | 16.9 |
| DAR | $H_M^{cl}$ | *131.4* | *140.4* | 149.3 | 158.3 | 160.4 | 161.3 | 166.2 | *171.1* | *175.9* | NA | NA | NA | NA |
| | $\Delta H_M^{ae}$ | -2.8 | -2.1 | -1.4 | -0.8 | 0.6 | 2.1 | 1.9 | *1.7* | *1.4* | NA | NA | NA | NA |
| FUA | $H_M^{cl}$ | *82.3* | *114.7* | 136.1 | 157.6 | 171.6 | *165.3* | 158.9 | 162.8 | *135.3* | 107.8 | 78.9 | *50* | 1521.2 |
| | $\Delta H_M^{ae}$ | 6.7 | *13.5* | 9.7 | 5.9 | 5.2 | *4.2* | 3.2 | 5.3 | *3* | 0.7 | 0.3 | *-0.2* | 57.4 |
| GOB | $H_M^{cl}$ | 249.9 | 208 | 210 | 174.2 | 156 | 136.1 | 146.7 | 168.2 | 193.8 | 225.3 | 247.8 | 258.9 | 2374.9 |
| | $\Delta H_M^{ae}$ | 3.1 | 1.9 | 0.5 | 0.4 | 0.4 | 0.2 | 1 | 3.6 | 4.8 | 3.9 | 2 | 1.5 | 23.5 |
| IZA | $H_M^{cl}$ | 143.8 | 153.2 | 207.1 | 235.7 | 280.5 | 280.7 | 283.5 | 256.6 | 213.2 | 168 | 131.8 | 126.3 | 2480.5 |
| | $\Delta H_M^{ae}$ | 0.8 | 0.1 | 0.6 | 1.2 | 0.9 | 2.3 | 6 | 4.7 | 0.8 | 0.3 | 0.1 | 0.2 | 18 |
| LRC | $H_M^{cl}$ | 73.5 | 85.1 | 132.5 | 173.2 | 192.3 | *185.8* | 179.3 | 180.2 | *146* | 111.8 | 78.7 | 59.8 | 1598.2 |
| | $\Delta H_M^{ae}$ | 0 | 0.3 | 1 | 1.2 | 3.2 | *3* | 2.8 | 2.9 | *1.9* | 0.8 | 0.1 | 0.3 | 17.4 |
| PAL | $H_M^{cl}$ | 9.9 | 47.6 | 85.4 | 135.7 | 164.3 | 174.5 | 183.6 | 149.4 | 111.4 | 72.8 | *34.3* | *22.1* | 1191 |
| | $\Delta H_M^{ae}$ | -0.9 | 1.1 | 3.1 | 3.5 | 1.6 | 0.7 | 2.4 | 1.7 | 0.4 | 1.3 | *2.1* | *0.6* | 17.6 |
| SXF | $H_M^{cl}$ | *45.8* | 81.4 | 117 | 158.7 | 190.1 | 208.9 | 219.1 | 167 | 139.6 | 97.3 | 64.6 | *31.8* | 1521.2 |
| | $\Delta H_M^{ae}$ | *-0.7* | -0.1 | 0.5 | 1.4 | 1.4 | 1.1 | 2.4 | 1.3 | 2.2 | 0.1 | -0.1 | *-0.2* | 9.2 |
| TAM | $H_M^{cl}$ | *111.5* | 137.4 | 163.3 | 160.6 | *151* | 141.4 | *158.7* | 176 | 147.8 | *141* | 134.3 | *122.9* | 1746 |
| | $\Delta H_M^{ae}$ | *-3.2* | 2.9 | 8.9 | 7.8 | *11.1* | 14.4 | *14.8* | 15.2 | 8.5 | *6.2* | 3.8 | *0.3* | 90.7 |
| TOR | $H_M^{cl}$ | -3.6 | 31.8 | 67.2 | 117.6 | 171.7 | *225.8* | 279.9 | NA | 69.9 | NA | NA | *-39* | NA |
| | $\Delta H_M^{ae}$ | -0.6 | -0.2 | 0.2 | 1.8 | 0.7 | *-0.5* | -1.6 | NA | 3 | NA | NA | *-1.1* | NA |
| XIA | $H_M^{cl}$ | 72.6 | 78.3 | 130 | 163.2 | 182 | *170.3* | 158.7 | 147.3 | 121.5 | 99.1 | 64.9 | 48.8 | 1436.7 |
| | $\Delta H_M^{ae}$ | 3.4 | 13.7 | 10.9 | 19.1 | 13.5 | *9.3* | 5.1 | 2.4 | 4.1 | 11.9 | 4.4 | 6.1 | 103.7 |

[37] Li, Z., Xia, X., Cribb, M., Mi, W., Holben, B., Wang, P., Chen, H., Tsay, S.C., Eck, T.F., Zhao, F. and Dutton, E.G., 2007. Aerosol optical properties and their radiative effects in northern China. *Journal of Geophysical Research: Atmospheres*, *112*(D22).

[38] Guiot, J. and Cramer, W., 2016. Climate change: The 2015 Paris Agreement thresholds and Mediterranean basin ecosystems. *Science*, *354*(6311), pp.465-468.

[39] Xia, X., 2014. A critical assessment of direct radiative effects of different aerosol types on surface global radiation and its components. *Journal of Quantitative Spectroscopy and Radiative Transfer*, *149*, pp.72-80.

[40] Christensen MW, Jones WK, Stier P. Aerosols enhance cloud lifetime and brightness along the stratus-to-cumulus transition. PNAS USA 2020;117(30):17591-17598.

[41] Khvorostyanov VI, Curry JA. Deliquescence and Efflorescence in Atmospheric Aerosols. Thermodynamics, Kinetics, and Microphysics of Clouds. Cambridge University Press, Cambridge, UK. 2014; pp.547-576.

[42] Stefu N, Paulescu M, Blaga R, Calinoiu D, Pop N, Boata R, Paulescu E. A theoretical framework for Ångström equation. Its virtues and liabilities in solar energy estimation. Energ Convers Manage 2016;112:236-245.

[43] Badescu V, Paulescu M. Statistical properties of the sunshine number illustrated with measurements from Timisoara (Romania). Atmos Res 2011;101(1-2):194-204.

[44] Paulescu M, Badescu V. New approach to measure the stability of the solar radiative regime. Theor Appl Climatol 2011;103(3):459-470.

[45] Gueymard, C.A. and Yang, D., 2020. Worldwide validation of CAMS and MERRA-2 reanalysis aerosol optical depth products using 15 years of AERONET observations. *Atmospheric Environment*, *225*, p.117216.

[46] Sun, X., Bright, J.M., Gueymard, C.A., Acord, B., Wang, P. and Engerer, N.A., 2019. Worldwide performance assessment of 75 global clear-sky irradiance models using principal component analysis. *Renewable and Sustainable Energy Reviews*, *111*, pp.550-570.

[47] Sun, X., Bright, J.M., Gueymard, C.A., Acord, B., Wang, P. and Engerer, N.A., 2019. Worldwide performance assessment of 75 global clear-sky irradiance models using principal component analysis. *Renewable and Sustainable Energy Reviews*, *111*, pp.550-570.

[48] Yang, D., 2021. Validation of the 5-min irradiance from the National Solar Radiation Database (NSRDB). *Journal of Renewable and Sustainable Energy*, *13*(1).
32

33[49] Monteiro, A., Basart, S., Kazadzis, S., Votsis, A., Gkikas, A., Vandenbussche, S., Tobias, A., Gama, C., García-Pando, C.P., Terradellas, E. and Notas, G., 2022. Multi-sectoral impact assessment of an extreme African dust episode in the Eastern Mediterranean in March 2018. *Science of The Total Environment*, *843*, p.156861.